# Transient interactions between bubbles and a high-speed cylinder in underwater launches: An experimental and numerical study

Sai Zhang,[1] Qihang Chen,[1] Chang Liu,[1] A-Man Zhang,[1] and Shuai Li[1*]

1. College of Shipbuilding Engineering, Harbin Engineering University, Harbin 150001, China

**Abstract**

The underwater launch of high-speed vehicles involves complex bubble–structure interactions, which are not currently well understood. In this study, two small-scale experiments are carried out involving transient bubble–cylinder interactions. We adopt the underwater electric discharge method to generate a high-pressure bubble that drives a cylinder to a maximum velocity of ~25 m/s within 1 ms. A tail bubble forms as the cylinder is ejected from the launch tube. Moreover, we observe a shoulder cavity around the head of the cylinder due to the pressure reduction in the flow. To better understand the complex interaction between bubbles and the high-speed cylinder, we use the boundary element method to establish a bubble–structure interaction model. Our numerical model reproduces the experimental observations quite well, including the cylinder motion and the transient evolution of the bubbles. Thereafter, a systematic study is carried out to reveal the dependence of the bubble–cylinder interactions on the initial pressure of the tail bubble $P_0$. We obtain a scaling law for the maximum velocity of the cylinder $v_\mathrm{m}$ with respect to $P_0$, namely, $v_\mathrm{m} \propto P_0^{0.45}$. The findings from this study may provide a reference for subsequent research into underwater launches.

---

\* Projects supported by the National Key R&D Program of China (2022YFC2803500), the Heilongjiang Provincial Natural Science Foundation of China (YQ2022E017), and the National Natural Science Foundation of China (52088102).

**Biography:** Sai Zhang (1998-), Female, Master Candidate, E-mail: 15645011104@163.com

**Corresponding author:** Shuai Li, E-mail: lishuai@hrbeu.edu.cn

# Introduction

The hydrodynamic characteristics associated with underwater launch technology include fluid–structure interactions, transient bubble dynamics, cavitation, etc. In recent years, underwater launch technology has received increasing attention from many countries. The high-pressure gas at the bottom of the launch tube pushes the vehicle upward. After the vehicle has completely left the tube, the high-pressure gas overflows and becomes attached to the bottom of the vehicle, forming a tail bubble[1, 2]. As the vehicle travels at high speed in the water, the shoulder pressure is lower than the saturated vapor pressure. Cavitation occurs in this low-pressure region and forms a shoulder bubble[3-5]. The bubble pulsation and jet may produce high levels of noise, change the forces acting on the vehicle, and affect the vehicle's trajectory. Therefore, studying the bubble–structure interactions in the underwater launch process is of great significance.

Numerical and experimental methods have been used to investigate the interactions between the vehicle and the shoulder bubble. Wu et al.[6] used the moving particle semi-implicit method to simulate the water-exit process. Fu et al.[7] found that the pulsation of the shoulder and tail bubbles decreased as the initial velocity increased, and obtained the relation between the shoulder bubble size and the initial vehicle velocity. Chen et al.[8] studied the details of the re-entrant jet produced by the ventilation shoulder bubble through decompression tank experiments and observed two types of the re-entry jet and four modes of development of ventilation shoulder bubbles' closure patterns. Lu et al.[9] adopted the matched asymptotic method to investigate the factors influencing the development characteristics of shoulder bubbles, including the pressure oscillations, head shape, and vehicle velocity. Zhao et al. [10] conducted an intensive experimental exploration of the effects of head shape on the flow characteristics and pressure load characteristics of the shoulder bubble at different stages. Chen et al.[11] used the commercial ANSYS FLUENT software to further simulate the evolution of the shoulder bubble during deceleration. They focused on the collapse of the shoulder bubble and investigated the influence of the angle of attack on the collapse process.

Several studies have examined the interactions between the tail bubble and the vehicle. Zahid et al.[12] simulated the evolution of the tail bubble with FLUENT and discussed the shock wave caused by its collapse. Ma et al.[13, 14] used FLUENT to establish a three-dimensional numerical model of the tail bubble, designed experiments to verify its effectiveness, and discussed the influence of the lateral velocity of the launcher. Dyment et al.[15] proposed an asymptotic model for the tail bubble prior to its collapse and designed an experimental device for underwater launch. After the first collapse of the tail bubble, wake vortices were generated as the vehicle moved upward. Xu et al.[16] used Star-CCM+ to study the evolution of wake vortices behind the vehicle. Cheng and Liu[17] proposed a gas–water flow coupling algorithm to solve the underwater ignition tail bubble problem. For the gas flow, the Euler equation was solved using the explicit non-oscillation and non-free-parameter dissipation scheme, while the water flow was solved from the Laplace equation based on potential theory and the boundary element method (BEM). Gao et al.[18] studied the dynamic characteristics of bubbles near the free surface.

As reviewed above, there have been some insightful works on the hydrodynamic characteristics of underwater launches using the finite volume method implemented in commercial software or experimental methods. Numerical methods provide specific information about the flow field, but are often computationally intensive. Experiments are realistic and intuitive, but are generally quite expensive. Recently, Cheng et al.[19] studied the tail bubble dynamics before pinch-off using a computationally simple BEM, and reported results very similar to those from CFD models. They explored the effects of various governing parameters on the tail bubble, including the vehicle velocity, water depth, initial gas volume, and initial gas pressure. However, the vehicle maintained a constant velocity in their simulations, which is effectively a one-way coupling model.

In this paper, we aim to identify more physical insights into the bubble–cylinder interactions that occur during the underwater launch process via both experimental and numerical approaches. First, we perform two small-scale experiments involving bubble–cylinder interactions and use a high-speed camera to capture the transient process. The underwater electric discharge method is used to generate a high-pressure bubble that drives a cylinder. A full interaction model is also proposed to model the bubble–cylinder interaction, in which the auxiliary function method is used to calculate the acceleration of the cylinder. The numerical results are found to be in good agreement with the experimental observations. Finally, we explore the dependence of the bubble–cylinder interaction on the initial pressure of the tail bubble.

# 1 Methodology

## 1.1 Experimental setup

A schematic of the experimental setup is shown in Fig. 1. The experimental equipment mainly includes a discharge box, copper wires, high-speed camera, light source, water tank, cylinder, and tube. The cylinder and the tube are made of acrylic. The experiments are performed in a cubic water tank (0.5×0.5×0.5 m). The acrylic rod is a cylinder (28.5 mm in height, 2.5 mm in radius, 0.85 g in weight). The launcher is an acrylic tube (30 mm in height, 4 mm in outer diameter, 2.8 mm in inner diameter). The electric discharge method[20-22] is adopted to produce high-pressure bubbles. The room temperature is around 20°C. The tank is filled with water up to the required height, and the cylinder is positioned in the center of the bottom of the tank. One end of each copper wire is bonded to the bottom of the tube, and the other end is connected to the discharge box. The cylinder is placed in the tube. The underwater spark discharge produces a high-pressure gas that pushes the cylinder upward. This process is captured by a high-speed camera with LED lights. In the high-speed movement of the cylinder, the shoulder bubble is formed by cavitation in the low-pressure area of the shoulder. The bottom gas spills out of the tube and attaches to the bottom of the cylinder with the movement of the cylinder, forming the tail bubble.

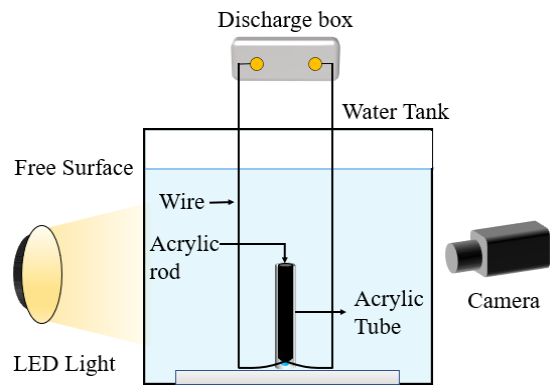

FIG. 1. Schematic diagram of the experimental equipment.

## 1.2 Numerical model

The high-pressure gas at the bottom of the tube pushes the cylinder out of the tube. The shoulder bubble is formed by cavitation at the shoulder of the cylinder[10]. The lifetime of the bubbles is quite short (~2.5 ms) and the Reynolds number can be estimated to be $O(10^5)$. Thus, the liquid viscosity can be ignored. The compressibility of the fluid can also be neglected because the characteristic velocity of the system is much smaller than the speed of sound in water. Therefore, the fluid is assumed to be inviscid, incompressible, and the flow is considered to be irrotational. Consider an axisymmetric configuration, the Laplace equation in a cylindrical coordinate system can be written as

$$\frac{\partial^2 \phi}{\partial z^2} + \frac{\partial^2 \phi}{\partial r^2} + \frac{1}{r}\frac{\partial \phi}{\partial r} = 0 \tag{1}$$

where $\phi$ is the velocity potential.

The bubble surface $S_b$ can be updated by

$$\frac{d\bm{x}}{dt} = \nabla \phi \quad \text{on} \quad S_b \tag{2}$$

where $\bm{x}$ denotes the field point. On the structural surfaces $S_s$, we have

$$\frac{\partial \phi}{\partial n} = \bm{u}_s \cdot \bm{n} \quad \text{on} \quad S_s \tag{3}$$

where $\bm{u}_s$ stands for the velocity vector of the solid structure. The pressure in the flow field satisfies

$$\frac{\partial \phi}{\partial t} = \frac{p_\infty - p}{\rho} - \frac{|\nabla \phi|^2}{2} - gz \tag{4}$$

where $p$ is the pressure, $\rho$ is the density of the fluid, $g$ is the gravitational acceleration and $p_\infty$ is the ambient pressure. The above equation can be written as

$$\frac{d\phi}{dt} = \frac{|\nabla \phi|^2}{2} + \frac{P_\infty}{\rho} - \frac{P}{\rho} - gz \tag{5}$$

In this study, we neglect the effect of surface tension because the Weber number is much greater than 1. Therefore, the water pressure at the surface of the tail bubble is approximately equal to the gas pressure in the tail bubble $p_b$, which can be derived from the adiabatic law[23] as

$$p = p_{\mathrm{b}} = p_{\mathrm{c}} + p_0 \left(\frac{V_0}{V}\right)^{\gamma} \tag{6}$$

where $p_{\mathrm{c}}$ is the vapor pressure, $V$ is the gas volume, $p_0$, $V_0$ denote the initial corresponding physical quantities of the bubble and $\gamma$ is the specific heat ratio. The shoulder bubble is a cavitation bubble, so its internal pressure $p_{\mathrm{s}}$ is set to the saturated vapor pressure.

An axisymmetric model[20-21, 24-25] is employed to simulate the bubble evolution. On the basis of Green's formula, the boundary integral equation on the boundaries can be derived from Eq. (1) in the form

$$2\pi\phi = \iint_S \left(\frac{\partial \phi}{\partial n} G - \phi \frac{\partial G}{\partial n}\right) \mathrm{d}S \tag{7}$$

where $S$ is the boundary surface, $G$ is the Green's function.

The hydrodynamic force on the cylinder is given by

$$\boldsymbol{F} = -\iint_{S_{\mathrm{s}}} \left(\phi_{\mathrm{t}} + \frac{1}{2}|\nabla \phi|^2 + \delta^2 z\right) \boldsymbol{n} \mathrm{d}S \tag{8}$$

According to Newton's second law, the motion of the cylinder is governed by

$$\boldsymbol{F} + \boldsymbol{F}_{\mathrm{e}} = M \cdot \dot{\boldsymbol{U}} \tag{9}$$

where $M$ is the mass of the cylinder, $\boldsymbol{F}_{\mathrm{e}}$ is the aggregation of other forces on the cylinder (such as gravity and the gas pressure at the bottom of the cylinder) and $\dot{\boldsymbol{U}}$ is the acceleration of the cylinder. On the wetted surface of the cylinder, we have

$$\frac{\partial \phi_{\mathrm{t}}}{\partial n} = \dot{\boldsymbol{U}} \cdot \boldsymbol{n} - \boldsymbol{U} \cdot \frac{\partial \nabla \phi}{\partial n} \tag{10}$$

where $\phi_{\mathrm{t}}$ can be decomposed into

$$\phi_{\mathrm{t}} = \eta + \kappa \tag{11}$$

On the surface of the bubbles, we have

$$\eta = 1 - \varepsilon \left(\frac{V_0}{V}\right)^{\kappa} - \frac{|\nabla \phi|^2}{2} - \delta^2 z, \quad \kappa = 0 \tag{12}$$

On the surface of the cylinder, the following holds

$$\frac{\partial \eta}{\partial n} = -\boldsymbol{U} \cdot \frac{\partial \nabla \phi}{\partial n}, \quad \frac{\partial \kappa}{\partial n} = \dot{\boldsymbol{U}} \cdot \boldsymbol{n} \tag{13}$$

The equation of $z$-direction motion of the suspended body is written as

$$F = -\iint_{S_s} \left( \eta + \frac{1}{2}|\nabla \phi|^2 + \delta^2 z \right) n_z \mathrm{d}S - \iint_{S_s} \kappa \cdot n_z \mathrm{d}S \tag{14}$$

The auxiliary functions $\beta$, $\chi$ are introduced to satisfy

$$\eta = \beta - \boldsymbol{U}\nabla\phi \tag{15}$$

$$\kappa = \dot{U}_z \chi \tag{16}$$

After considering the gravity of the cylinder

$$(M + P_b + \iint_{S_s} \chi \cdot n_z \mathrm{d}S) \cdot \dot{U}_z = -\iint_{S_s} \left( \eta + \frac{1}{2}|\nabla\phi|^2 + \delta^2 z \right) n_z \mathrm{d}S - \delta^2 M \tag{17}$$

The acceleration of the cylinder $\dot{U}_z$ can be calculated using Eq. (17). From the acceleration, the velocity can easily be obtained. Finally, through the known velocity potential of the bubble and the normal velocity of the cylinder, the normal velocity of the bubbles and the velocity potential of the cylinder are obtained. The location of the bubbles is then updated. The time step is determined by

$$\Delta t = \frac{\Delta \phi}{\max \left| \frac{P_\infty}{\rho} + \frac{1}{2}|\nabla\phi|^2 - gz - \frac{P}{\rho} \right|} \tag{18}$$

where $\Delta\phi$ is a constant value.

## 2 The verification and convergence analysis of BEM model

This section describes two experiments involving transient bubble–cylinder interactions. The results of the numerical simulations are compared with the experimental data. Sensitivity and convergence analyses are performed to further verify our numerical model.

### 2.1 Comparison of experimental and numerical results

For the simulations, the parameters are set to be the same as in the experiments. We define $t = 0$ as the moment when the spark bubble is formed. The water depth at the tube mouth is 0.3 m, the water density is 1025 kg/m³, and the acceleration of gravity is 9.8 m/s². The origin of the coordinate system is set at the middle of the tube mouth, with the radial direction denoted by $r$ and the axial direction denoted by $z$. The lapped copper wire discharges continuously produce gas for a period of time. The gas generation process is difficult to simulate, so a unidirectional interaction model is first established. The cylinder velocity observed in the experiment is used as the cylinder velocity in the numerical simulations. It is difficult to simulate the shoulder cavity formed by the cavitation of the shoulder low-pressure region with the BEM, so the calculations start after the cylinder has moved a small distance upward. The initial node layout of the model is shown in Fig. 2(a) and the node layout after the cylinder has completely left the tube is shown in Fig. 2(b).

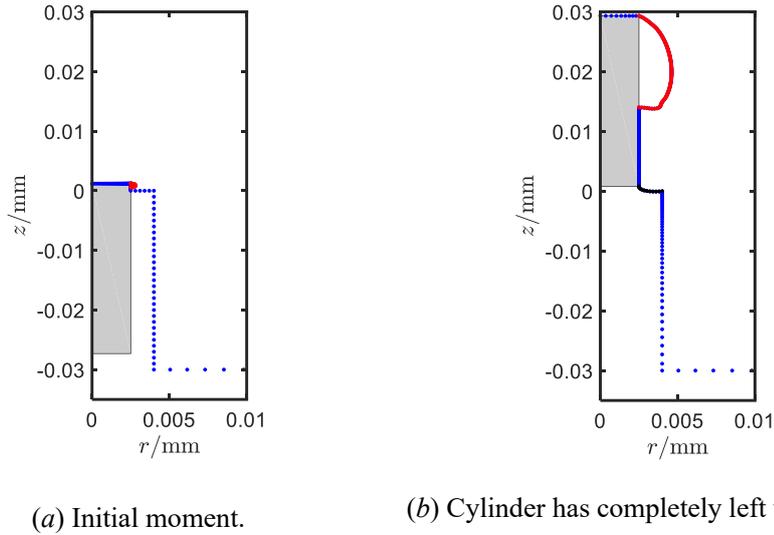

(a) Initial moment.　　　(b) Cylinder has completely left the tube.

FIG. 2. BEM node layout. The gray rectangle represents the cylinder, the blue points mark the structure, the red points mark the shoulder cavity, and the black points mark the tail cavity.

We select $t = 0.2$ ms as the physical time for the simulation initialization. At this time, the cylinder velocity and displacement are set according to the experimental data. However, the spark bubbles produce a strong light source, which prevents the high-speed camera from accurately measuring the position of the cylinder. When the burning of the copper wire ends and the strong light weakens, the displacement of the cylinder in the experiment can be measured and the speed can be calculated. To better describe the physical process in the initial stage of the experiment, it is assumed that the motion state of the cylinder can be divided into two periods of uniform acceleration when the light is strong. The displacement and velocity of the cylinder at the initial stage can be obtained by ignoring the change of acceleration of the cylinder[20]. The initial volume and pressure of the spark bubble are difficult to measure. After fixing the initial volume of the tail gas in the BEM model, different initial pressures are applied until the trajectory of the tail bubble in the model is consistent with that in the experiment. The experimental results are compared with the numerical results to verify the validity of the BEM model for simulating the shoulder and tail bubbles. Figures 3 and 4 compare the numerical and experimental data produced by discharge voltages of 900 V and 1100 V, respectively. At the initial time of the BEM model, the 900-V discharge has produced a cylinder displacement of 1.12 mm and velocity of 10.96 m/s, a tail gas volume of 27.49 mm$^3$, and an initial pressure of 5.59 MPa. In the case of the 1100-V discharge, the cylinder's initial displacement and velocity are 1.19 mm and 11.91 m/s, the tail gas volume is 29.33 mm$^3$, and the initial pressure is 9.37 MPa.

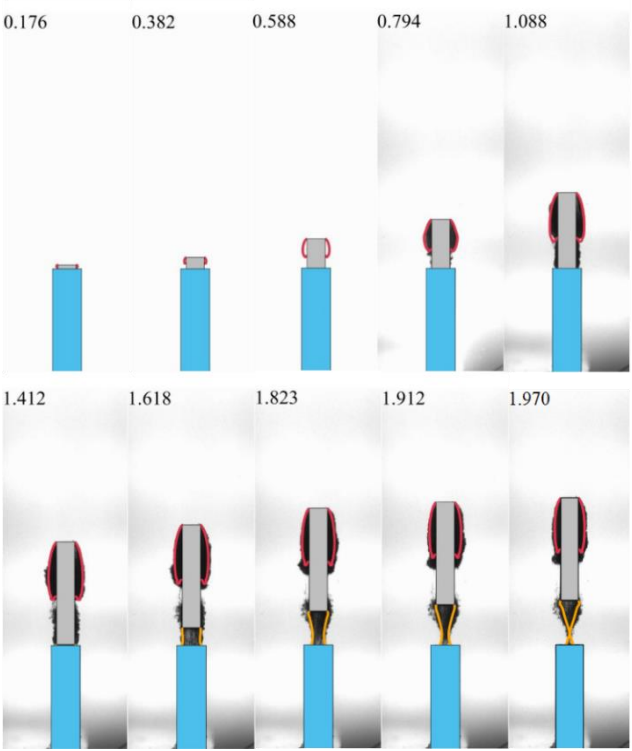

FIG. 3. One-to-one comparison between experimental results and numerical simulations obtained from BEM. The red lines denote the profiles of the shoulder bubble and the yellow lines represent the tail bubble. The gray rectangles denote the cylinder. The discharge voltage is 900 V in the experiment.

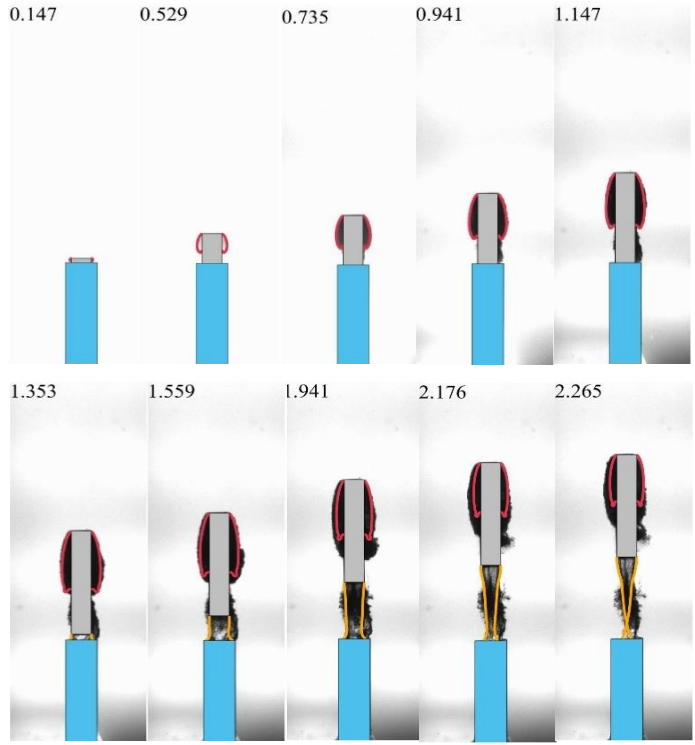

FIG. 4. One-to-one comparison between experimental results and numerical simulations obtained from BEM. The discharge voltage is 1100 V in the experiment.

Two lapped copper wires at the tube bottom produce bubbles through joule heating. The strong light generated during this process prevents the camera from shooting clearly. The high-pressure gas pushes the cylinder to move upward. A shoulder cavity is formed around the head of the high-pressure cylinder due to the reduced pressure in the flow. The shoulder bubble expands with increasing cylinder velocity. After the cylinder has completely left the tube, the bottom gas spills out of the tube, attaches to the tail of the cylinder, and continues to stretch axially. As the tail bubble volume increases, the pressure of the tail bubble and the thrust received by the cylinder decrease. Once the pressure in the tail cavity is lower than the ambient pressure, the radial expansion of the tail bubble slows down. Once the maximum bubble radius has been attained, the tail bubble starts to collapse, eventually pinching off above the tube opening. As observed in Figs. 3 and 4, the shoulder bubble in the BEM results is slightly smaller than that observed in the experiment. This may be because the dissolved non-condensable gas in the surrounding water enters the shoulder bubble in the experiment, while the gas core in the surrounding water is ignored in the numerical simulations. In the experiment, before the cylinder has completely left the tube, some gas escapes through the gap between the cylinder and the tube. With the upward movement of the cylinder, the spillover gas fuses with the tail bubble. The spillover gas is not considered in the BEM model, resulting in some differences in the tail cavity. In general, the experiments and the numerical simulations are not affected by the small amount of spillover gas. The BEM results are in good agreement with the trend of the experimental results.

Next, the BEM was used to establish a full interaction model for numerical calculations, and the parameters were set to be the same as in the experiment. The initial pressure was calculated within the range of $[8.0\times10^6, 1.4\times10^7]$ Pa, and the cylinder speed was compared with the experimental observations. The comparison results are presented in Fig. 5. The triangles indicate the experimental results and the dashed lines are the time curves of the cylinder when the pressure is at the range boundaries. The solid lines are the two boundaries of the comparison between the calculated BEM results and the experimental values. In the experiment, the gas volume continued to increase for a period of time after discharge, while the gas volume of the BEM model was fixed. This discrepancy produces some calculation errors. Overall, the results in Fig. 5 demonstrate the validity of the full interaction model.

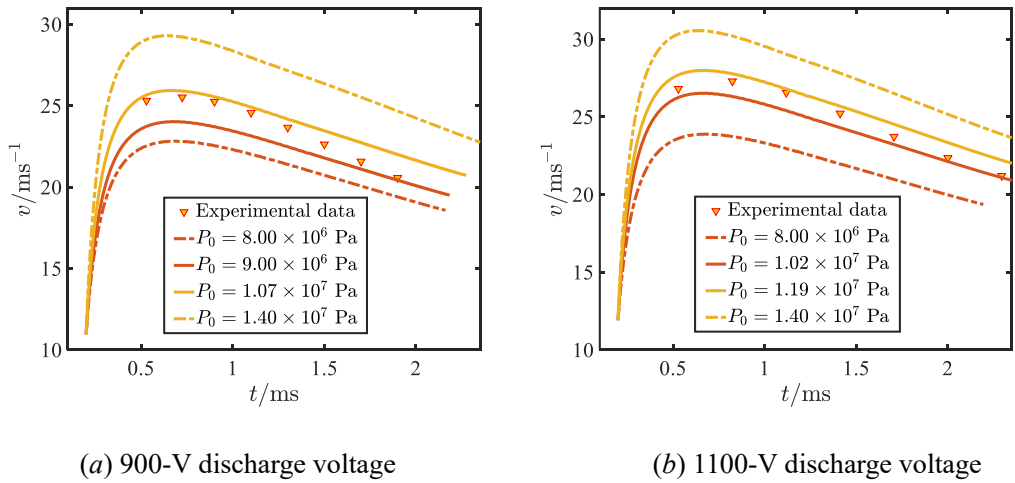

(*a*) 900-V discharge voltage   (*b*) 1100-V discharge voltage

FIG. 5. Comparison of cylinder velocities between the BEM model and experiments.

## 2.2 Sensitivity and convergence analysis

The generation of the shoulder bubble is difficult to simulate by the BEM, so the numerical model calculations start after the cylinder has moved a short distance. At the initial moment, we set a semicircle (in the axisymmetric model) to represent the shoulder cavity, and the diameter of the semicircle is the shoulder bubble length $l_0$. To verify the independence of the initial shoulder bubble size, the initial pressure of the tail bubble was set as 9.37 MPa and the value of $l_0$ was varied from 0.2–0.8 mm, with all other parameters unchanged. The variation of the shoulder bubble length $l$ with time is shown in Fig. 6. The results indicate that the length of the initial shoulder bubble does not affect the accuracy of the calculation results. Thus, a value of $l_0 = 0.6$ mm was selected for all subsequent calculations.

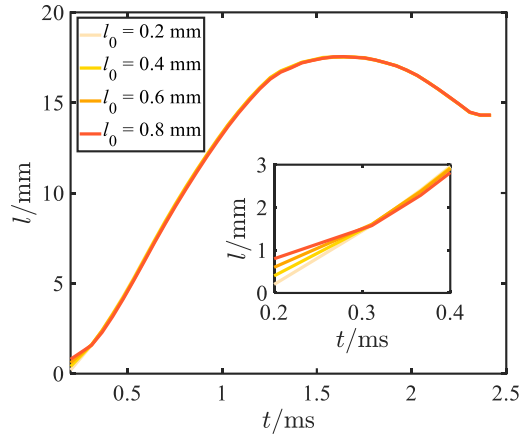

FIG. 6. Independence of the shoulder bubble length with respect to initial bubble size.

In the initial setup, all mesh sizes were adapted to the mesh size of the shoulder bubble. We examined different mesh sizes while leaving the other parameters unchanged. The convergence of the model was verified by comparing the calculated results. The initial mesh size of the shoulder bubble was varied by changing the mesh number $n$ from 20–80. Comparing the cylinder velocity curves in Fig. 7, the calculation results with $n = 60$ and 80 tend to be consistent, verifying the convergence of the interaction model. To balance the calculation efficiency with the accuracy of the model, subsequent calculations used $n = 60$.

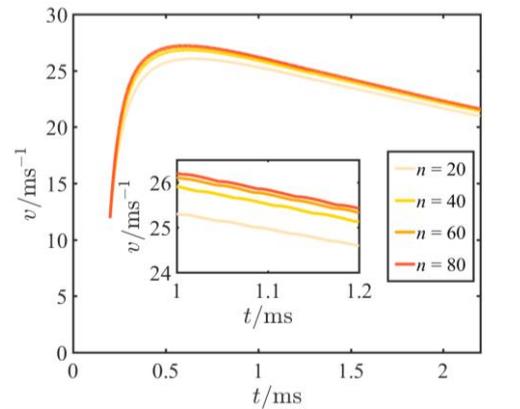

FIG. 7. Convergence study of the mesh number of the initial shoulder bubble.

# 3 Discussion

The verified numerical model was used to calculate the pressure and velocity fields after the cylinder exited the tube in the 1100-V experiment, as shown in Fig. 8. The left side of each panel shows the velocity contours and the right side of each panel shows the pressure contours. At the head of the cylinder, the fluid velocity is large and there is a high-pressure zone. A high-velocity re-entrant jet appears at the wake of the shoulder bubble, and a high-pressure zone appears below this area of high velocity. As the cylinder moves upward, the tail bubble stretches in the axial direction. After expanding to its maximum extent in the radial direction, the tail bubble starts to collapse radially and gradually accelerates. Finally, the tail bubble pinches off near the mouth of the tube.

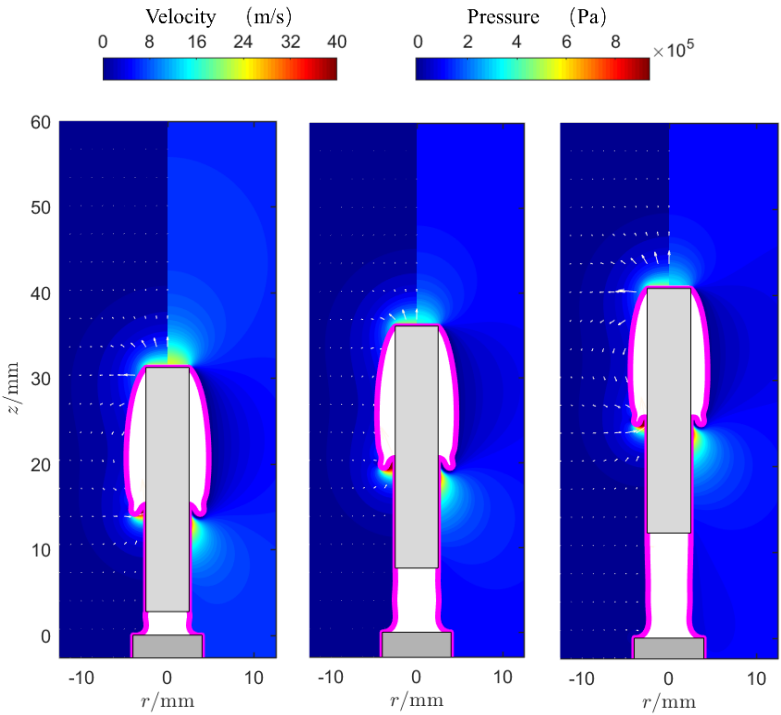

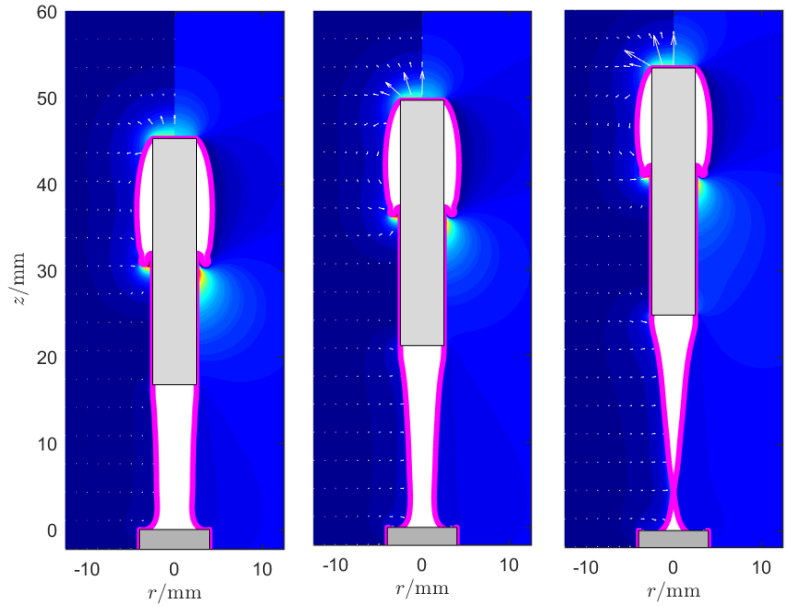

FIG. 8. Velocity and pressure fields calculated using the BEM model for the 1100-V experiment. The left side of each panel shows the speed and the right side shows the pressure. Arrows denote the velocity field.

Referring to the experimental data described above, the calculation of the BEM model starts at $t = 0.2$ ms, when the cylinder displacement is 1.1 mm and the velocity is 11 m/s (unless otherwise stated). The full interaction BEM model was used to numerically simulate the underwater launch process with different initial tail bubble pressures. The maximum velocity $v_m$ of the cylinder and the maximum length $l_m$ of the shoulder bubble at different initial pressures are shown in Fig. 9. The maximum velocity of the cylinder and the maximum length of the shoulder bubble increase as the initial pressure rises. Figure 10 shows the relation between the logarithm of the initial pressure and the maximum velocity. The data points lie on a straight line with a slope of 0.45, i.e., $v_m \propto P_0^{0.45}$. Initial pressures of $1.10 \times 10^7$ Pa, $1.87 \times 10^7$ Pa, and $2.56 \times 10^7$ Pa were considered in this study. The variation of the velocity and acceleration curves with time are shown in Fig. 11. The cylinder acceleration is greatest at the beginning of the calculation, with acceleration up to the order of $10^5$ m/s². The acceleration drops sharply to negative values within 0.3 ms. The cylinder velocity rises rapidly to its maximum value and then begins to decay. In the 1100-V experiment, the cylinder displacement is 7.1 mm at $t = 0.47$ ms. Assuming uniform acceleration in this period, the cylinder acceleration is $6.43 \times 10^4$ m/s². Because the actual acceleration decreases rapidly, the maximum acceleration will be greater than $6.43 \times 10^4$ m/s². This is consistent with the above conclusion that the maximum acceleration can be up to the order of $10^5$ m/s$^2$. The cylinder is accelerated upward by the high-pressure gas. As the volume of the tail bubble increases, the pressure inside the tail bubble and the cylinder acceleration decrease. After reaching its maximum velocity, the cylinder starts to decelerate. In the range of pressures studied herein, a higher initial tail bubble pressure produces a higher maximum cylinder velocity and faster pinch-off.

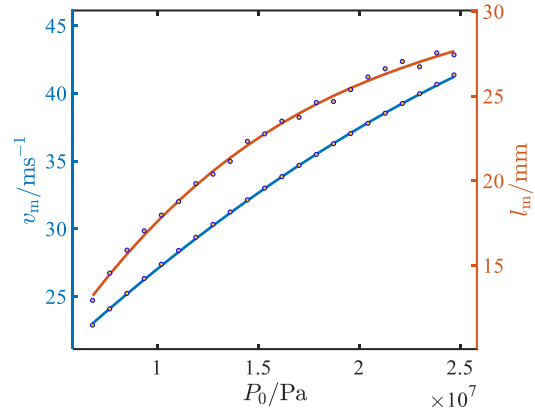

FIG. 9. Maximum cylinder velocity and maximum shoulder bubble length at different initial pressures. Blue denotes the fitted curve of the maximum velocity. Red denotes the fitted curve of the maximum shoulder bubble length.

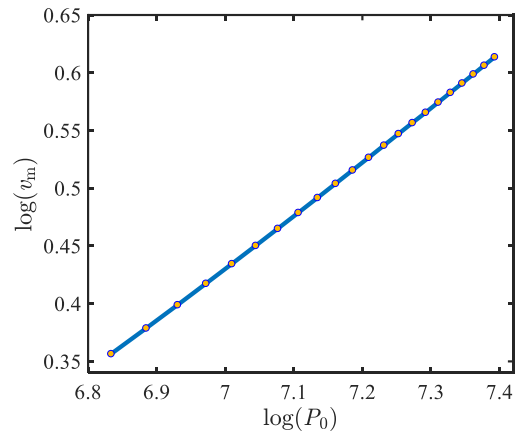

FIG. 10. Logarithmic relationship between initial pressure and maximum cylinder velocity.

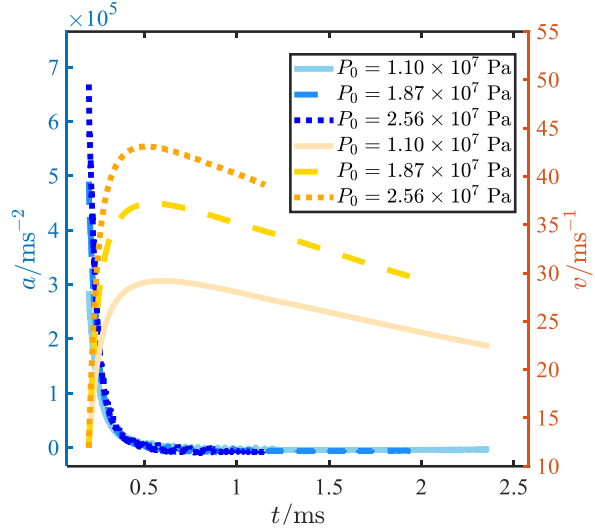

FIG. 11. Variation of velocity and acceleration with time.

Figure 12 plots the velocity at different moments as the horizontal coordinate and the length $l$ and thickness $W$ of the shoulder bubble as vertical coordinates. As the cylinder velocity increases monotonically to its maximum and then decreases monotonically, the data points can be connected to form a smooth curve with respect to time. The length and thickness of the shoulder bubble gradually increase as the cylinder accelerates. When the maximum cylinder velocity is reached, the length and thickness of the shoulder bubble continue to grow. The shoulder bubble reaches its maximum size after the cylinder has been decelerating for a period of time. The time at which the length and thickness reach their maximum values is not consistent. The length and thickness then start to decrease as the cylinder slows down.

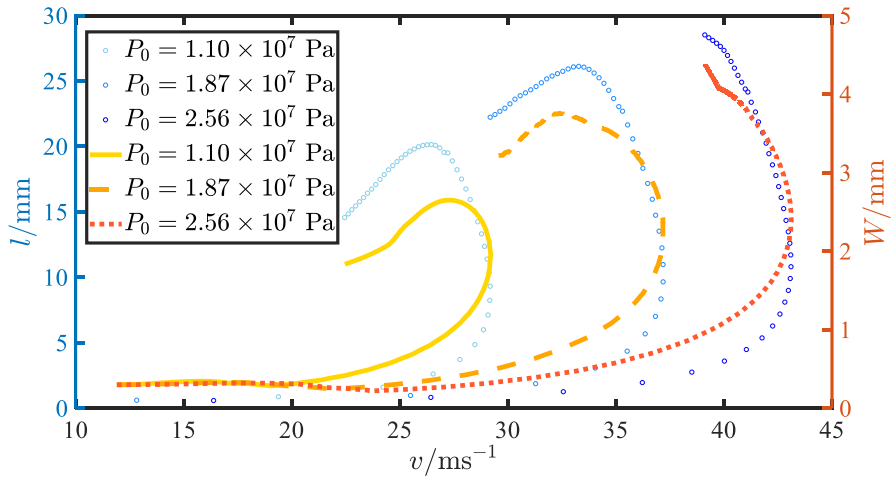

FIG. 12. Variation in shoulder bubble length and thickness with velocity. The length of the shoulder bubble is denoted by the blue dots and the thickness is denoted by the red/orange dots/lines.

# 4 Conclusions

This paper has described the use of experimental and numerical methods to study the transient interactions between bubbles and a high-speed cylinder in underwater launches. In the experiments, an underwater electric discharge was used to generate a high-pressure bubble that drives the cylinder. A fully coupled bubble–structure interaction model was established using the BEM, enabling us to gain deeper insights into underwater launches. Our model reproduces the experimental observations quite well, including the cylinder motion and the transient evolution of bubbles. Finally, we explored the dependence of the bubble–cylinder interactions on the initial pressure of the tail bubble. The main findings are as follows:

1) A scaling law was obtained for the maximum velocity of the cylinder $v_m$ with respect to $P_0$, namely, $v_m \propto P_0^{0.45}$.

2) The maximum acceleration of the cylinder can reach the order of $10^5$ m/s$^2$ at the initial moment, before rapidly decreasing to a negative value.

3) The shoulder cavitation bubble grows quickly as the cylinder is driven upward by the high-pressure bubble at its base. In our numerical simulations, an annular high-pressure region was generated behind the shoulder bubble, leading to the formation of a re-entrant jet.

4) The tail bubble is stretched by the cylinder in the vertical direction. After the tail bubble expands to its maximum radius, the gas pressure inside the tail bubble is much lower than the ambient pressure. Thereafter, the tail bubble collapses and finally pinches off.

## Acknowledgments

This work is supported by the National Key R&D Program of China (2022YFC2803500), the Heilongjiang Provincial Natural Science Foundation of China (YQ2022E017), and the National Natural Science Foundation of China (52088102).

## Statements and Declarations

**Conflict of Interest:** The authors declare that they have no conflict of interest.
**Ethical approval:** This article does not contain any studies with human participants or animals performed by any of the authors.

**Informed consent:** Informed consent was obtained from all individual participants included in the study.

## References

[1] Chen S. R., Shi Y., Pan G., and Gao S. Experimental Research on Cavitation Evolution and Movement Characteristics of the Projectile during Vertical Launching. *Journal of Marine Science and Engineering*, 2021, 9(12): 1359.

[2] Wang Y. W., Liao L. J., Du T. Z., Huang C. G., Liu Y. B., Fang X., and Liang N. G. A study on the collapse of cavitation bubbles surrounding the underwater-launched projectile and its fluid–structure coupling effects. *Ocean Engineering*, 2014, 84: 228-236.

[3] Wang Y. W., Huang C. G., Fang X., Wu X. C., and Du T. Z. On the internal collapse phenomenon at the closure of cavitation bubbles in a deceleration process of underwater vertical launching. *Applied Ocean Research*, 2016, 56: 157-165.

[4] Yu X. X., Huang C. G., Du T. Z., Liao L. Z., Wu X. C., Zheng Z., and Wang Y. W. Study of Characteristics of Cloud Cavity Around Axisymmetric Projectile by Large Eddy Simulation. *Journal of Fluids Engineering*, 2014, 136(5): 051303.

[5] Cao J. Y., Lu C. J., Chen Y., Chen X., and Li J. Research on the Base Cavity of a Sub-Launched Projectile. *Journal of Hydrodynamics*, 2012, 24(2): 244-249.

[6] Wu Q. R., Wang L., Xie Y. H., Du Z. H., and Yu H. M. Numerical Simulation of the water-exit process of the missile based on Moving Particle Semi-Implicit method. *Journal of Physics: Conference Series*, 2019, 1300(1): 012064.

[7] Fu G. Q., Zhao J. L., Sun L. P., and Lu Y. Experimental Investigation of the Characteristics of an Artificial Cavity During the Water-Exit of a Slender Body. *Journal of Marine Science and Application*, 2018, 17(4): 578-584.

[8] Chen X. B., Xiao W., Gong R. Y., Yao X. L., and Hu S. F.. Experimental investigation of ventilation bubble dynamics around a vertically moving cylinder under reduced ambient pressure. *Fluid Dynamics Research*, 2022, 54(1): 015512.

[9] Lu J. X., Wang C., Wei Y. J., Sun T. Z., Liu F., and Xu H. Experimental and theoretical investigation of the cavity dynamics of underwater launched projectiles. *Ocean Engineering*, 2022, 254: 111291.

[10] Zhao Q. K., Chen T., Xiao W., Chen X., B., Yao X. L., and Wang W. P. Research on the characteristics of cavitation flow and pressure load during vertical water exit of different head-shaped vehicles. *Ocean Engineering*, 2022, 265: 112663.

[11] Chen Y., Lu C. J., Chen X., and Cao J. Y. Numerical investigation on the cavitation collapse regime around the submerged vehicles navigating with deceleration. *European Journal of Mechanics - B/Fluids*, 2015, 49: 153-170.

[12] Zahid M. Z., Nadeem M., and Ismail M. Numerical Study of Submarine Launched Underwater Vehicle. *2020 17th International Bhurban Conference on Applied Sciences and Technology (IBCAST) IEEE*. 2020: 472-476.

[13] Ma G. H., Chen F., Yu J. Y., and Jiang S. Effect of a pressure-equalizing film on the trajectory and attitude robustness of an underwater vehicle considering the uncertainty of the platform velocity. *Engineering Applications of Computational Fluid Mechanics*, 2018, 12(1): 824-838.

[14] Ma G. H., Chen F., Yu J. Y., and Wang K. Numerical Investigation of Trajectory and Attitude

Robustness of an Underwater Vehicle Considering the Uncertainty of Platform Velocity and Yaw Angle. *Journal of Fluids Engineering*, 2019, 141(2): 021106.

[15] Dyment A., Flodrops J. P., Paquet J. B., Dupuis D., and Marchand D. Gaseous cavity at the base of an underwater projectile. *Aerospace science and technology*, 1998, 2(8): 489-504.

[16] Xu H., Wei Y. J., Wang C., and Lu J. X. On wake vortex encounter of axial-symmetric projectiles launched successively underwater. *Ocean Engineering*, 2019, 189: 106382.

[17] Cheng Y. S., and Liu H. Mathematical modeling of fluid flows for underwater missile launch. *Journal of Hydrodynamics*, 2006, 18(S1): 481-486.

[18] Gao S., Shi Y., Pan G., and Quan X. B. A study on the performance of the cavitating flow structure and load characteristics of the vehicle launched underwater. *Physics of Fluids*, 2022, 34(12): 125108.

[19] Cheng S. H., Quan X. B., Zhang S., Zhang T. Y., and Li S. Modeling tail bubble dynamics during the launch of an underwater vehicle using the boundary element method. *Journal of Hydrodynamics*, 2022, 34(3): 434-443.

[20] Li S., Zhang A. M., Wang S. P., and Han R. Transient interaction between a particle and an attached bubble with an application to cavitation in silt-laden flow. *Physics of Fluids*, 2018, 30(8): 082111.

[21] Han R., Zhang A. M., Tan S. C., and Li S. Interaction of cavitation bubbles with the interface of two immiscible fluids on multiple time scales. *Journal of Fluid Mechanics*, 2021, 932: A8.

[22] Zhang A. M., Li S. M., Cui P., Li S., and Liu Y. L. A unified theory for bubble dynamics. *Physics of Fluids*, 2023, 35(3), 033323.

[23] Best J. P. The formation of toroidal bubbles upon the collapse of transient cavities. *Journal of Fluid Mechanics*, 2006, 251: 79-107.

[24] Xue Y. Z., Cui B., and Ni B. Y. Numerical study on the vertical motion of underwater vehicle with air bubbles attached in a gravity field. *Ocean Engineering*, 2016, 118: 58-67.

[25] Cui R. N., Li S., Wang S. P., and Zhang A. M. Pulsating bubbles dynamics near a concave surface. *Ocean Engineering*, 2022, 250: 110989.

[26] Liu Y. L., Wang S. P., and Zhang A. M. Interaction between bubble and air-backed plate with circular hole. *Physics of Fluids*, 2016, 28(6): 062105.